\documentclass[a4paper, 12pt, oneside]{article}
\usepackage[cp1251]{inputenc}
\usepackage[english]{babel}
\usepackage{graphics, amsfonts, amsmath, bm, amssymb, array}
\usepackage[dvips]{graphicx}
\usepackage[flushleft,small]{caption2}

\newcommand{\diag}{\rm \diag\, }
\newcommand{\intl}{\int\limits}

\makeatother {

\begin{document}
\thispagestyle{empty} \large

\renewcommand{\refname}{\begin{center}{\bf REFERENCES}\end{center}}

\begin{center}
{\bf Surface Plasmons in Thin Metallic Films}
\end{center}

\begin{center}
  \bf  A. V. Latyshev\footnote{$avlatyshev@mail.ru$} and
  A. A. Yushkanov\footnote{$yushkanov@inbox.ru$}
\end{center}\medskip

\begin{center}
{\it Faculty of Physics and Mathematics,\\ Moscow State Regional
University,  105005,\\ Moscow, Radio st., 10--A}
\end{center}\medskip

\begin{abstract}
For the first time it is shown that for thin metallic films
thickness of which not exceed thickness of skin -- layer, the
problem of description of surface plasma oscillations
allows analytical solution by arbitrary ratio between length of
electrons free path and thickness of a film.
The dependance of frequency surface plasma oscillations on
wave number is carry out.
\medskip

\medskip

\end{abstract}

\begin{center}\bf
  Introduction
\end{center}

Electromagnetic properties of metal films already in a current
long time are a subject of steadfast attention
\cite{F69} -- \cite{F09}.
Recently special interest involves in itself
a problem about surface plasma oscillations  \cite{Economou}
-- \cite{Arx2009-2}.
It is connected as with theoretical interest to this
problem, and with numerous practical appendices as well.
Thus the majority of researches is founded on
the description of properties of films with use of methods
macroscopical electrodynamics.
For thin films such approach is inadequate, as for the description of films
in the thickness of an order and less than length of mean free path
of electrons macroscopical electrodynamics is inapplicable.
The electrons scattering on a surface demands kinetic consideration.
It is serious complicates the problem.

In the present work it is shown that for thin films, a thickness of
which does not exceed a thickness of a skin -- layer, the problem
of description of surface plasma oscillations
allows the analytical solution  by arbitrary ratio between length of
mean free path of electrons and thickness of a film..

Let's notice, that the most part of reasonings carrying out below is
fair for more general case of conducting medium  (in particular,
semi-conductor) films.\medskip
\vskip5mm

\begin{center}
{\bf Statement problem}
\end{center}

Let's consider a thin metal film.

We take Cartesian coordinate system with origin of coordinates on
one of the surfaces of a slab, with axes $x$, directed deep into the
slab and perpendicularly to the surface of a film.
The axes $z$ we will direct along a direction of propagation
of the surface electromagnetic wave.
We will notice, that in this case a magnetic field
is directed along an axis $y$.

At such choice of system of coordinates the electric field
vector and magnetic field vector have the following structure
$$
\mathbf{E}
=\{E_x(x,z,t),0,E_z(x,z,t)\}, \quad
\mathbf{H}
=\{0,H_y(x,z,t),0\}.
$$

The origin of coordinates we will place on the bottom plane limiting a film.
Let's designate a thickness of a film through $d$.

Out of the film the electromagnetic field is described by the equations
$$
\dfrac{1}{c^2}\dfrac{\partial^2 {\bf E}}{\partial t^2}-
\Delta {\bf E}=0
$$
and
$$
\dfrac{1}{c^2}\dfrac{\partial^2 {\bf H}}{\partial t^2}-\Delta{\bf
H}=0.
$$

Here $c$ is the velocity of light, $\Delta$ is the Laplace operator.

The solution of these equations decreasing on infinity, looks like
$$
{\bf E}=\left\{\begin{array}{l}{\bf E}_1e^{-i\omega t+\alpha x+ikz},\qquad
x<0, \\
{\bf E}_2e^{-i\omega t+\alpha (d-x)+ikz},\qquad x>d,
\end{array}\right.
\eqno{(1a)}
$$
and
$$
{\bf H}=\left\{\begin{array}{l}{\bf H}_1e^{-i\omega t+\alpha x+ikz},\qquad
x<0, \\
{\bf H}_2e^{-i\omega t+\alpha(d-x)+ikz},\qquad x>d.
\end{array}\right.
\eqno{(1b)}
$$

Here $\omega$ is the frequency of wave, $k$ is the number wave,
damping parameter $\alpha$ is connected with these quantities
by relation
$$
\alpha=\sqrt{k^2-\dfrac{\omega^2}{c^2}},
\eqno{(2)}
$$
$\mathbf{E}_j$ and $\mathbf{H}_j \;(j=1,2)$ are constant amplitudies.

Further components of intensity vectors electric and
magnetic fields we search in the following form
$$
E_x(x,z,t)=E_x(x)e^{-i\omega t+ikz},\quad
E_z(x,z,t)=E_z(x)e^{-i\omega t+ikz},
$$
and
$$
H_y(x,z,t)=H_y(x)e^{-i\omega t+ikz}.
$$

Then behaviour of electric and magnetic
fields of the wave in the film is described by the following system
the differential equations \cite{K}
$$
\left\{\begin{array}{l}
\dfrac{dE_z}{dx}-ik E_x+\dfrac{i\omega}{c}H_y=0, \\ \\
\dfrac{i\omega}{c}E_x-ik H_y=\dfrac{4\pi}{c}j_x,\\ \\
\dfrac{dH_y}{dx}+\dfrac{i\omega}{c}E_z=\dfrac{4\pi}{c}j_z.
\end{array}\right.
\eqno{(3)}
$$

Here  $\,\bf j$ is the current density.

The equations (3) are satisfied and out of the film  under the
condition ${\bf j}=0$.

Impedance on the bottom surface of the layer (film) then
is defined as follows
$$
Z=\dfrac{E_z(-0)}{H_y(-0)}.
\eqno{(4)}
$$

We consider in the given work the case, when
$z$ -- component of electric field has the antisymmetric
configuration concerning of the film middle.
Then \cite {F66} $y$ -- component of magnetic field and $x $ --
component of electric field have the symmetric configuration
concerning of the film middle. Thus
$$
H_y(0)=H_y(d), \qquad E_x(0)=E_x(d),\qquad E_z(0)=-E_z(d).
\eqno{(5)}
$$

It is required to find a spatial dispersion of the surface
plasmon, i.e. to find dependence of frequency of oscillations  own
mode of system (3) on quantity of the wave vector
$\omega=\omega(k)$. \\

\begin{center}
{\bf Surface plasmon}
\end{center}

Let's consider the case when the width of a layer $d$ is less than depth
skin -- layer $\delta$. We will notice that depth skin -- layer essentially
depends on frequency of radiation, monotonously decreasing in process
of growth last. The quntity $\delta$ accepts the minimal value in
so-called infra-red case \cite{Landau10}
$$
\delta_0=\dfrac{c}{\omega_p},
$$
where $\omega_p$ is the plasma frequency.

For typical metals \cite{Landau10} $\delta_0\sim 10^{-5}$ cm.

Thus for the films which thickness $d$ is less
$\delta_0$, our assumption holds for any frequencies.

Quantities $H_y$ and $E_z$ a little vary on distances smaller
than depths of skin -- layer.
Therefore at performance of the given assumption ($d <\delta_0$)
this field will vary a little in the layer.

Let's consider the first of conditions (5) $H_y(0)=H_y(d)$.
Because of this condition it is possible to accept, that the
quantity $H_y$ is constant in the layer. Change of quantity $z$ --
projection of electric field on
the thickness of the layer can define from the first equation of system (3)
$$
E_z(d)-E_z(0)=-\dfrac{i\omega}{c}d H_y+ik \intl_0^d E_x dx.
\eqno{(6)}
$$

From the second equation of system (3) taking into account a
non--flowing condition of a current
through boundary of the film and a continuity condition of
electric and magnetic fields follows, that on film border holds
the relation
$$
E_x(0)=E_x(d)=\dfrac{ck}{\omega}H_y.
\eqno{(7)}
$$

The integral entering into the relation (6) is proportional to value of
quantity of normal to the surface of component of electric field on
surfaces, and consequently to the quantity $H_y$. Therefore it is natural
to enter proportionality coefficient
$$
G=\dfrac{1}{E_x(0)}\intl_0^d E_x(x)dx.
$$

For the case $kl\ll 1$  the quantity $G$ can be calculated
from the problem about behaviour of a plasma layer in variable
electric field, perpendicular to the surface layer \cite {LY2008}.

Taking into account (7) this coefficient  will be copied in the form
$$
G=\dfrac{1}{H_y\Big(\dfrac{ck}{\omega}\Big)}\intl_0^d E_x(x)dx.
\eqno{(8)}
$$

Hence, expression (6) with use (8) then can be written down as
$$
E_z(d)-E_z(0)=ikdH_y\Big(1-\dfrac{ck}{\omega}G\Big).
$$

Considering antisymmetric character $z$ -- projection of
electric field $E_z$ in this case we receive
$$
E_z(0)=ik\dfrac{d}{2}H_y\Big(1-\dfrac{ck}{\omega}G\Big).
\eqno{(9)}
$$

According to (9) for an impedance (4) we have
$$
Z=ik\dfrac{d}{2}\Big(1-\dfrac{ck}{\omega}G\Big).
\eqno{(10)}
$$

From the third equation of system (3)
taking into account relations (1) we receive the following
connection between $y$ -- projection of magnetic field and $z$ --
projection of electric field in the immediate vicinity
from the bottom surface of the layer and out of it (when $j_z=0$)
$$
\alpha H_y(0)=-\dfrac{i\omega}{c}E_z(0).
$$

From here we receive following expression for the surface impedance
$$
Z=\dfrac{i\alpha c}{\omega}.
\eqno{(11)}
$$

Equating expressions (10) and (11), we receive
$$
\dfrac{\alpha c}{\omega}=\dfrac{kd}{2}\Big(1-\dfrac{ck}{\omega}G\Big).
\eqno{(12)}
$$

The expression (11) can be transformed according the relation (2)
to the form
$$
\sqrt{k^2-\dfrac{\omega^2}{c^2}}=\dfrac{\omega
kd}{2c}\Big(1-G\dfrac{ck}{\omega}\Big).
\eqno{(13)}
$$

The equation (13) is the dispersion equation, from solution of which
we find the connection $\omega=\omega(k)$.

In general case the function $G$, entered by the relation (8),
is the function of two variables: $G=G (\omega,k)$.
Therefore the dispersion equation
(13) represents the difficult transcendental equation.

Let's consider further a case of low frequencies. We take such
frequencies that essentially are less than frequency
a volume plasma resonance of metal.
In this case  $|G|\ll 1$. Then the dispersion equation (13) is possible
to transform to the following kind
$$
(ck)^2-\omega^2=\dfrac{\omega^2k^2d^2}{4}.
$$

The solution of this equation we will write as
$$
\omega^2=\dfrac{4(ck)^2}{4+k^2d^2},
$$
hence
$$
\omega(k)=\dfrac{ck}{\sqrt{1+\Big(k\dfrac{d}{2}\Big)^2}}.
$$

At small values of a wave vector $k$ when $kd\ll 1$, from here we receive
$$
\omega(k)=ck\Big(1-\dfrac{k^2d^2}{8}\Big).
$$

\begin{center}
{\bf Conclusion}
\end{center}

In the present work the dispersion relation  for surface plasmon
is deduced. We consider the case of an antisymmetric configuration
of $z$ -- component of the electric field, directed lengthways
propagation of an electromagnetic wave, and symmetric
$y$ -- component of a magnetic field and $x$ -- component of electric
field.

\end{document}